\documentclass[letterpaper,twocolumn,english,prl]{revtex4}

\usepackage[T1]{fontenc}
\usepackage[utf8x]{inputenc}
\usepackage{amsmath}
\usepackage{graphicx}

\makeatletter

\@ifundefined{textcolor}{}
{%
 \definecolor{BLACK}{gray}{0}
 \definecolor{WHITE}{gray}{1}
 \definecolor{RED}{rgb}{1,0,0}
 \definecolor{GREEN}{rgb}{0,1,0}
 \definecolor{BLUE}{rgb}{0,0,1}
 \definecolor{CYAN}{cmyk}{1,0,0,0}
 \definecolor{MAGENTA}{cmyk}{0,1,0,0}
 \definecolor{YELLOW}{cmyk}{0,0,1,0}
}

\@ifundefined{definecolor}
 {\usepackage{color}}{}

\makeatother

\usepackage{babel}
\begin{document}

\title{First Passage Under Restart }

\author{{\normalsize{}Arnab Pal$^{*}$ \& Shlomi Reuveni$^{\dagger}$}\\
{\normalsize{}~}}

\affiliation{\noindent \textit{$^{*}$Schulich Faculty of Chemistry, Technion—Israel
Institute of Technology, Technion City, Haifa 32000, Israel.}}

\affiliation{\noindent \textit{$^{\dagger}$Department of Systems Biology, Harvard
Medical School, Boston, Massachusetts 02115, USA.}}
\begin{abstract}
{\normalsize{}First passage under restart has recently emerged as
a conceptual framework suitable for the description of a wide range
of phenomena, but the endless variety of ways in which restart mechanisms
and first passage processes mix and match hindered the identification
of unifying principles and general truths. Hope that these exist came
from a recently discovered universality displayed by processes under
optimal, constant rate, restart—but extensions and generalizations
proved challenging as they marry arbitrarily complex processes and
restart mechanisms. To address this challenge, we develop a generic
approach to first passage under restart. Key features of diffusion
under restart—the ultimate poster boy for this wide and diverse class
of problems—are then shown to be completely universal. }{\normalsize \par}
\end{abstract}
\maketitle
A myriad of basic questions and a wide array of applications have
turned first passage time (FPT) processes into a long standing focal
point of scientific interest \cite{Redner,Metzler}. These processes
were studied extensively, e.g. in the context of nonequilibrium systems
\cite{Schehr}, but despite many years of study paramount discoveries
are still being made and exciting applications continue to be found.
Recently, several groups have observed that any FPT process imaginable
can become subject to restart, i.e., can be stopped and started anew
(Fig. 1). This observation has opened a rapidly moving theoretical
research front \cite{Restart1,Restart2,Restart3,Restart4,Restart5,Restart6,Restart7,Restart8,Restart9,Restart10,Restart11,Restart12,Restart13,Restart14,Restart15,Restart16}
and applications to search problems \cite{Restart-Search1,Restart-Search2,Restart-Search3},
the optimization of randomized computer algorithms \cite{Restart in CS-1,Restart in CS-2,Restart in CS-3,Restart in CS-4,Restart in CS-5,Restart in CS-6},
and in the field of biophysics \cite{Restart-Biophysics1,Restart-Biophysics2},
have further propelled its expansion. Universality has always been
considered a holy grail of the physical sciences and novel revelations
concerning universality in FPT processes have recently taken center
stage and attracted considerable attention \cite{Condamin,Benichou,Chupeau}.
In contrast, not a lot is known in general about the problem of first
passage under restart (FPUR).

\begin{figure}
\noindent \begin{centering}
\includegraphics[scale=0.2]{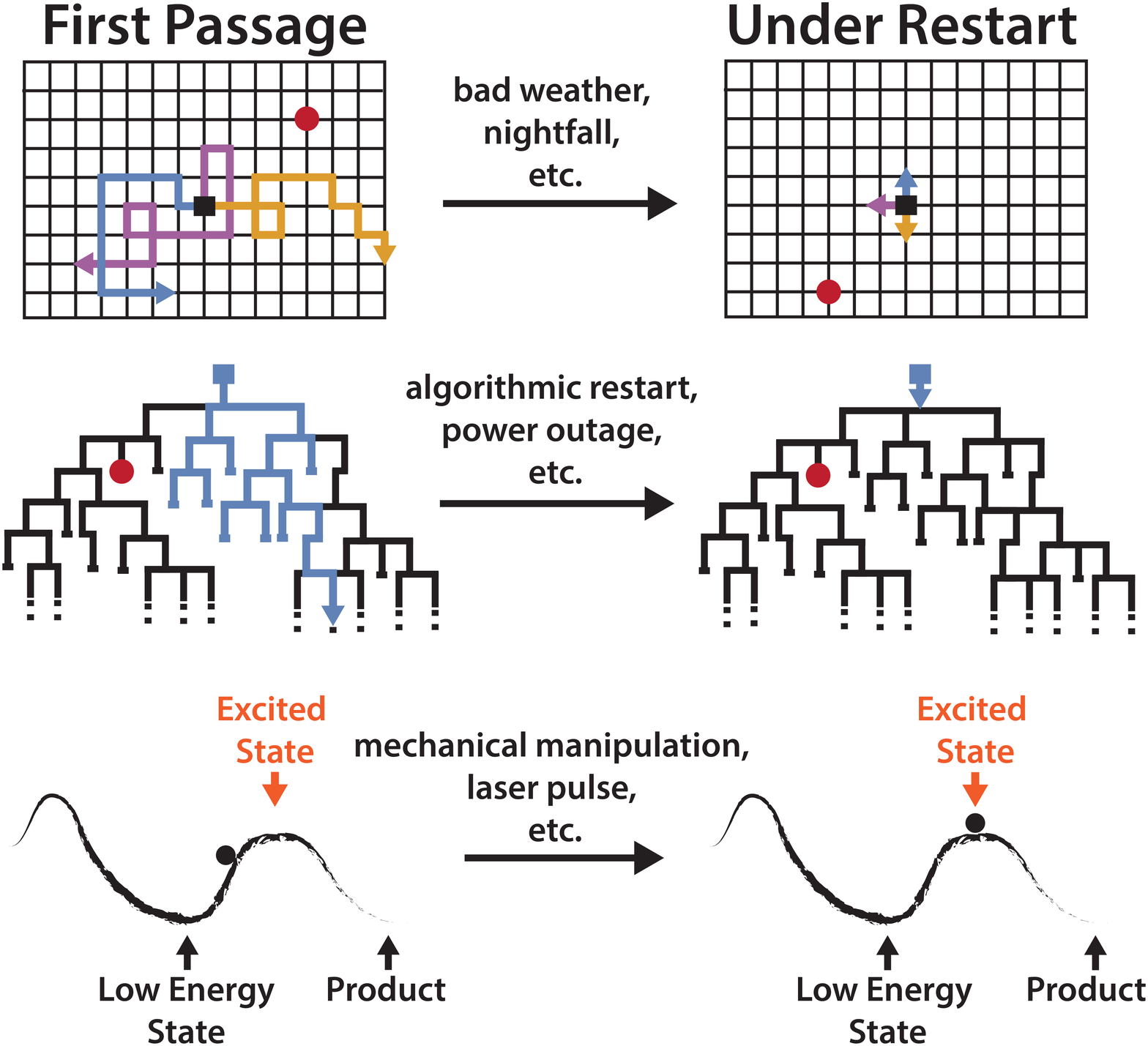}
\par\end{centering}
\caption{\textbf{Top.} Bad weather could force a team of searchers to temporally
cease their efforts and return to base. By the time search is renewed
the target may have relocated and search must thus start from scratch.
\textbf{Middle.} A computer algorithm operates as a black box which
randomly scans a tree of possibilities in search of a solution. Chance
may send the algorithm down the wrong path but programmed restart
could help rescue the search. \textbf{Bottom.} A molecule that was
previously prepared at an excited state decays to a low energy state.
A pulse of laser could bring the molecule back to its excited state
and restart a chemical or physical reaction. This time, the desired
product may be formed. }
\end{figure}
Diffusion with resetting to the origin is a quintessential example
of FPUR \cite{Restart2}. In this problem, a particle undergoes diffusion
but from time to time is also taken and returned to the place from
where it started its motion (reset or restart). In addition, at some
distance away from the origin a target awaits and one is interested
in the time it takes the particle to first get to the target, i.e.,
in its distribution and corresponding moments. This problem was first
studied with restart rates that are constant in time and the surprise
came from the fact that restart was able to expedite search and that
a carefully chosen (optimal) restart rate could minimize the mean
FPT to the target. Further down the road, other restart mechanisms
were also studied \cite{Restart9,Restart10,Restart13,Restart16} and
it was shown that these may under- or over-preform when compared to
restart at a constant rate \cite{Restart9,Restart16}. 

Each variant above carried with it some unique and intriguing features,
but exhausting the vast combinatorial space of process-restart pairs—one
problem at a time—is virtually impossible. Indeed, restart processes
may take different shapes and forms and the effect they have on FPT
processes other than diffusion \cite{Other than diffusion1,Other than diffusion2,Other than diffusion3,Other than diffusion4,Other than diffusion5,Other than diffusion6,Other than diffusion7,Other than diffusion8}
is also of interest. Moreover, it is often the case—in real life scenarios—that
the process under consideration, the restart mechanism that accompanies
it, or both are poorly specified or even completely unknown. General
approaches, better suited to deal with partial and missing information
and with the need to generalize from specific examples, could then
become handy.

Recently, two attempts to unify treatment were made. In \cite{Restart15},
an approach suitable to the description of a generic FPT process under
constant rate restart was presented. The approach was utilized to
show that when restart is optimal—the relative fluctuation in the
FPT of the restarted process is always unity. This result holds true
regardless of the underlying process, be it diffusion or other, but
is no longer valid for time dependent restart rates as these were
not covered by the approach to begin with. Restart rates with arbitrary
time dependence were considered in \cite{Restart9}, but analysis
there was limited to diffusion and did not cover other FPT processes.
Here, we will be interested in merging the two approaches in attempt
to get the best of both worlds. To this end, we consider a generic
FPT process that has further become subject to a generic restart mechanism.
This setting is extremely general and captures, as special cases,
the overwhelming majority of models that have already appeared in
the literature. We analyze this scheme to attain, and concisely describe,
several broad scope results which unravel universal features of this
wide class of problems. In what follows, we use $f_{Z}(t)$, $\left\langle Z\right\rangle $,\textbf{
}$\sigma^{2}\left(Z\right)$ and $\tilde{Z}(s)\equiv\left\langle e^{-sZ}\right\rangle $
to denote, respectively, the probability density function, expectation,
variance, and Laplace transform of a real-valued random variable $Z$. 

\textbf{Mean FPT under restart.} Consider a generic process that starts
at time zero and, if allowed to take place without interruptions,
ends after a random time $T$. The process is, however, restarted
at some random time $R$. Thus, if the process is completed prior
to restart the story there ends. Otherwise, the process will start
from scratch and begin completely anew. This procedure repeats itself
until the process reaches completion. Denoting the random completion
time of the restarted process by $T_{R}$ it can be seen that

\noindent 
\begin{equation}
\begin{array}{l}
T_{R}=\left\{ \begin{array}{lll}
T &  & \text{if }T<R\text{ }\\
 & \text{ \ \ }\\
R+T_{R}^{\prime} &  & \text{if }R\leq T\text{ ,}
\end{array}\right.\text{ }\end{array}\label{1}
\end{equation}
where $T_{R}^{\prime}$ is an independent and identically distributed
copy of $T_{R}$. 

A scheme similar to the one described in Eq. (\ref{1}) was analyzed
in \cite{Restart15}. There, no assumptions were made on the distribution
of the time $T$ which governs the completion of the underlying process,
but the restart time $R$ was assumed to be exponentially distributed
with rate parameter $r$. This means that restart is conducted at
a constant rate $r$, i.e., that for any given time point the probability
that restart will occur at the next infinitesimal time interval $dt$
is $rdt$. Here, we relax this assumption allowing for generally distributed
restart times or, equivalently, for restart rates with arbitrary time
dependence. Letting $r(t)$ denote the restart rate at time $t$,
we note that the two perspectives are related via (Fig. 2) 
\begin{equation}
\mathrm{Pr}(R\leq t)=1-exp\left(-\int_{0}^{t}r(x)dx\right)\,,\label{2}
\end{equation}
where $Pr(R\leq t)$ is the probability that $R\leq t$ \cite{Reliability}.
\begin{figure}
\noindent \begin{centering}
\includegraphics[scale=0.35]{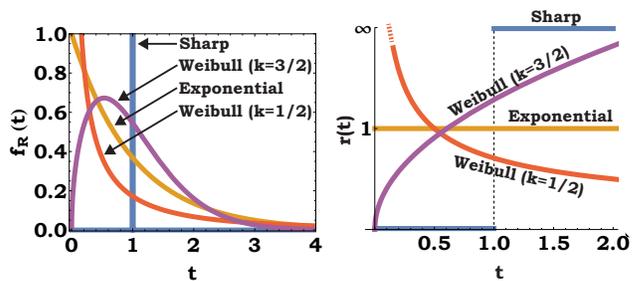}
\par\end{centering}
\caption{A few examples of restart time distributions (left) and the restart
rates they induce (right). Below $\delta(x)$ is the Dirac delta function,
$\varGamma(x)$ is the Gamma function, and $\left\langle R\right\rangle =1$
in all plots: (i) Deterministic (sharp) restart $f_{R}(t)=\delta\left(t-\left\langle R\right\rangle \right)$.
Restart rate jumps abruptly from zero to infinity at $t=\left\langle R\right\rangle $;
(ii) Exponentially distributed restart $f_{R}(t)=\left\langle R\right\rangle ^{-1}e^{-t/\left\langle R\right\rangle }$.
Restart rate is constant: $r(t)=1/\left\langle R\right\rangle $;
(iii \& iv) Weibull distributed restart $f_{R}(t)=\frac{k}{\lambda}\left(\frac{t}{\lambda}\right)^{k-1}e^{-\left(t/\lambda\right)^{k}}$.
Restart rate is given by $r(t)=kt^{k-1}/\lambda^{k}$ and could monotonically
decrease (e.g. $k=1/2$, $\lambda=\left\langle R\right\rangle /2$)
or increase (e.g. $k=3/2$, $\lambda=\left\langle R\right\rangle /\Gamma(5/3)$)
with time.}
\end{figure}

Equation (\ref{1}) could be used to provide a simple formula for
the mean FPT of a stochastic process under restart. Indeed, noting
that it can also be written as $T_{R}=min(T,R)+I\{R\leq T\}T_{R}^{\prime}\,,$
where $min(T,R)$ is the minimum of $T$ and $R$ and $I\{R\leq T\}$
is an indicator random variable which takes the value one when $R\leq T$
and zero otherwise, we take expectations to find 
\begin{equation}
\left\langle T_{R}\right\rangle =\frac{\left\langle min(T,R)\right\rangle }{Pr(T<R)}\,.\label{3}
\end{equation}
The right hand side of Eq. (\ref{3}) can then be computed given the
distributions of $T$ and $R$ if one also recalls that the cumulative
distribution function of $min(T,R)$ is given by $Pr(min(T,R)\leq t)=1-Pr(T>t)Pr(R>t).$ 

A hallmark of restart is its ability to minimize (optimize) mean FPTs.
For example, when the restart rate $r(t)=r$ is constant it is straight
forward to show that Eq. (\ref{3}) reduces to $\left\langle T_{R}\right\rangle =\left(1-\tilde{T}(r)\right)/\left(r\tilde{T}(r)\right),$
where $\tilde{T}(r)$ is the Laplace transform of $T$ evaluated at
$r$. One could then seek an optimal rate $r^{*}$ which brings $\left\langle T_{R}\right\rangle $
to a minimum, derive general conditions for this rate to be strictly
larger than zero, and further discuss universal properties of the
optimal rate itself \cite{Restart12,Restart-Biophysics1}. Clearly,
this line of inquiry is not limited to the case of exponentially distributed
restart times and could also be extended to other parametric distributions.
Various optimization questions could then be addressed directly, but
we would now like to consider a broader optimization question. Specifically,
we ask if within the vast space of stochastic restart strategies,
and irrespective of the underlying process being restarted, there
is a single winning strategy that could not be beat?

\textbf{Sharp restart is a dominant strategy}. Consider a particle
``searching'' for a stationary target via one dimensional diffusion.
The particle starts at the origin, the initial distance between the
particle and the target is $L$, and the diffusion coefficient of
the particle is $D$. Denoting the particle's FPT to the target with
$T$, the latter is known to come from the Lévy-Smirnov distribution
$f_{T}(t)=\sqrt{L^{2}/4D\pi t^{3}}e^{-L^{2}/4Dt}$ \cite{Redner}.
Considering the same problem under restart, we take $D=1/2$ and $L=1,$
and utilize Eq. (\ref{3}) to plot $\left\langle T_{R}\right\rangle $
as a function of $\left\langle R\right\rangle $ for various restart
time distributions (Fig. 3). As can be seen, a minimum of $\left\langle T_{R}\right\rangle $
is always attained and while the values taken by the different minima
and their positions clearly depend on the distribution of the restart
time—it is sharp restart that attains the lowest of minima. A similar
observation was made in the past and it was consequently conjectured
that in the case of diffusion mediated search sharp restart is the
optimal restart strategy \cite{Restart9,Restart16}. Strikingly, this
is also true in general.

Consider, for the sake of simplicity, a random restart time $R$ characterized
by a proper density $f_{R}(t)$ and note that $\left\langle min(T,R)\right\rangle =\int_{0}^{\infty}f_{R}(t)\left\langle min(T,R)|R=t\right\rangle dt=\int_{0}^{\infty}f_{R}(t)\left\langle min(T,t)\right\rangle dt\,,$
which then implies $\left\langle T_{R}\right\rangle =\int_{0}^{\infty}\frac{f_{R}(t)Pr(T<t)}{Pr(T<R)}\frac{\left\langle min(T,t)\right\rangle }{Pr(T<t)}dt$.
However, $\int_{0}^{\infty}\frac{f_{R}(t)Pr(T<t)dt}{Pr(T<R)}=1,$
and $\left\langle min(T,t)\right\rangle /Pr(T<t)$ is simply the mean
completion time of a process that is restarted sharply after $t$
units of time. Thus, if there exists some $t^{*}$ such that $\frac{\left\langle min(T,t^{*})\right\rangle }{Pr(T<t^{*})}\leq\frac{\left\langle min(T,t)\right\rangle }{Pr(T<t)}$
for all $t\geq0$—sharp restart at $t^{*}$ will also beat any random
restart time that is governed by a proper density. Moreover, the law
of total expectation implies $\left\langle min(T,R)\right\rangle =\left\langle \left\langle min(T,R)|R\right\rangle _{T}\right\rangle _{R}$
and steps similar to those taken above assert that (SI) 
\begin{equation}
\frac{\left\langle min(T,t^{*})\right\rangle }{Pr(T<t^{*})}\leq\frac{\left\langle min(T,R)\right\rangle }{Pr(T<R)}\,,\label{4}
\end{equation}
for any random restart time $R$ regardless of its distribution. Equation
(\ref{4}) thus asserts that sharp restart is optimal among all possible
stochastic restart strategies in continuous time, and we refer the
reader to Luby \emph{et.al. }for complementary, algorithm oriented,
discussion on the discrete time case \cite{Restart in CS-1}.
\begin{figure}
\noindent \begin{centering}
\includegraphics[scale=0.46]{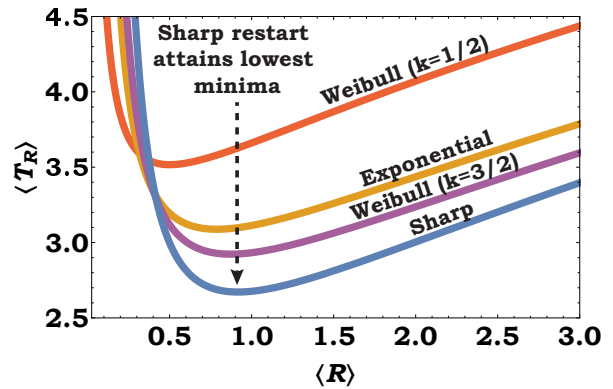}
\par\end{centering}
\caption{Mean FPT for diffusion mediated search with restart vs. the mean restart
time, for various restart time distributions taken from Fig. 2. }
\end{figure}
\begin{figure}
\noindent \begin{centering}
\includegraphics[scale=0.17]{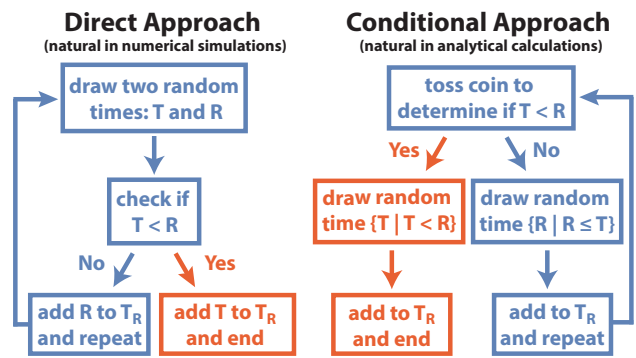}
\par\end{centering}
\caption{Two approaches to first passage under restart.}
\end{figure}

\textbf{Distribution of FPT under restart. }So far, we have only been
concerned with the \emph{mean} FPT of a restarted process but will
now move on to discuss the full distribution of $T_{R}$. The scheme
described in Eq. (\ref{1}) suggests a direct approach for numerical
simulation of FPUR (Fig. 4 left). In this approach, one draws two
random times from the distributions of $T$ and $R$, and only then—based
on the outcome of that draw—decides which of the two, restart or completion,
happened first. An equivalent approach would operate in reversed order.
A coin with probability $Pr(T<R)$ will first be tossed to determine
if completion preceded restart (or vice versa) and only then, given
that information, the appropriate—conditional—random time will be
drawn (Fig. 4 right). This approach is somewhat awkward and indirect
for the purpose of numerical simulations, but is actually quite natural
when coming to compute expectations and Laplace transforms where one
usually starts by conditioning on the occurrence of an event of interest.
Indeed, analytical formulas could be simplified with the aid of two
auxiliary random variables: $R_{min}\equiv\{R\,|\,R=min(R,T)\}$ and
$T_{min}\equiv\{T\,|\,T=min(R,T)\}$. In words, $R_{min}$ is the
random restart time given that restart occurred prior to completion,
and $T_{min}$ is defined in a similar manner. Conditioning on whether
$T<R$ and applying the law of total expectation to $\tilde{T}_{R}(s)=\left\langle e^{-sT_{R}}\right\rangle $,
we obtain (SI) 
\begin{equation}
\tilde{T}_{R}(s)=\frac{Pr(T<R)\tilde{T}_{min}(s)}{1-Pr(R\leq T)\tilde{R}_{min}(s)}\,.\label{5}
\end{equation}

Equation (\ref{5}) allows one to explicitly compute the distribution
of $T_{R}$ in Laplace space. For example, when $T$ and $R$ are
correspondingly governed by probability densities $f_{T}(t)$ and
$f_{R}(t)$, we have $Pr(T<R)=\int_{0}^{\infty}f_{T}(t)\left(\int_{t}^{\infty}f_{R}(t')dt'\right)dt$
and the probability densities governing $T_{min}$ and $R_{min}$
are similarly given by 
\begin{equation}
\begin{array}{l}
\begin{array}{c}
f_{T_{min}}(t)=f_{T}(t)\int_{t}^{\infty}f_{R}(t')dt'/Pr(T<R)\,,\\
\\
f_{R_{min}}(t)=f_{R}(t)\int_{t}^{\infty}f_{T}(t')dt'/Pr(R\leq T)\,.
\end{array}\text{ }\end{array}\label{6}
\end{equation}
Plugging in concrete probability distributions explicit formulas can
be obtained, e.g. for exponentially distributed restart $f_{R}(t)=re^{-rt}$
and one could readily show that $\tilde{T}_{R}(s)=\tilde{T}(s+r)/\left(\frac{s}{s+r}+\frac{r}{s+r}\tilde{T}(s+r)\right)$
(SI) as was previously obtained in \cite{Restart15} by other means. 

\textbf{Fluctuations in FPT under optimal sharp restart obey a universal
inequality. }Given Eq. (\ref{5}), one could utilize the known relation
between moments and the Laplace transform \cite{Other than diffusion7}
to find (SI) 
\begin{equation}
\left\langle T_{R}^{2}\right\rangle =\frac{\left\langle min(T,R)^{2}\right\rangle }{Pr(T<R)}+\frac{2Pr(R\leq T)\left\langle R_{min}\right\rangle \left\langle min(T,R)\right\rangle }{Pr(T<R)^{2}}\,.\label{7}
\end{equation}
A special case of this result was used to show that the relative fluctuation,
$\sigma\left(T_{R}\right)/\left\langle T_{R}\right\rangle $, is always
unity when a process is restarted at a constant rate $r^{*}>0$ which
brings $\left\langle T_{R}\right\rangle $ to a minimum. Optimal \emph{sharp}
\emph{restart} could lower the mean FPT, $\left\langle T_{R}\right\rangle $,
well below the value it attains for optimal \emph{constant rate restart}
but unless the resulting fold reduction is also matched or exceeded
by a fold reduction in $\sigma\left(T_{R}\right)$—the relative fluctuation
in the FPT would surely increase. It is thus possible that the ability
of the sharp restart strategy to attain lower mean FPTs comes at the
expense of higher relative fluctuations—and hence greater uncertainty—in
the FPT itself. However, when Eq. (\ref{7}) was utilized to examine
diffusion and other case studies (Fig. 5), we consistently found 
\begin{equation}
\sigma\left(T_{t^{*}}\right)/\left\langle T_{t^{*}}\right\rangle \leq1\,,\label{8}
\end{equation}
for the relative fluctuation at the optimal restart time $t^{*}$. 

Equation (\ref{8}) is universal. To see this, we assume by contradiction
that there exists a FPT process for which $\sigma\left(T_{t^{*}}\right)/\left\langle T_{t^{*}}\right\rangle >1$;
and consider a restart strategy $R_{mix}$ in which this process is
restarted at a low constant rate $r\ll1$ in addition to being sharply
restarted whenever a time $t^{*}$ passes from the previous restart
(or start) epoch. Applying this restart strategy is equivalent to
augmenting the process under sharp restart with an additional restart
mechanism that restarts it with rate $r$. However, if the relative
fluctuation in the FPT of a process is larger than unity—restart at
a low constant rate will surely lower its mean FPT (and vice versa).
This is true regardless of the underlying process, and can be seen
by examining $\left\langle T_{R}\right\rangle $ for general $T$,
and $R$ which is exponentially distributed with rate $r$ (see formula
below Eq. (\ref{3})). Utilizing the moment representation of the
Laplace transform, one can then show that $[d\left\langle T_{R}\right\rangle /dr]|_{r=0}<0$
whenever $\sigma\left(T\right)/\left\langle T\right\rangle >1$ (SI).
Denoting the mean FPT under $R_{mix}$ by $\left\langle T_{R_{mix}}\right\rangle $,
and letting $T_{t^{*}}$ take $T$'s place above, it follows that
$\left\langle T_{t^{*}}\right\rangle >\left\langle T_{R_{mix}}\right\rangle $.
We have thus found a non-sharp restart strategy which lowers the mean
FPT beyond that attained for optimal sharp restart. However, this
finding must be false as it stands in contradiction to the proven
dominance of optimal sharp restart (discussion above), and Eq. (\ref{8})
then follows immediately. More generally, an equation similar to Eq.
(\ref{8}) must hold for every restart strategy $R$ which attains
a FPT that cannot be lowered further by introducing an additional
restart rate $r\ll1$, and Eqs. (\ref{3}) and (\ref{7}) could then
be utilized to comprehensively characterize this set of strategies
(SI) 
\begin{equation}
\sigma\left(T_{R}\right)/\left\langle T_{R}\right\rangle \leq1\Longleftrightarrow\left\langle T_{min}\right\rangle \geq\frac{1}{2}\frac{\left\langle min(T,R)^{2}\right\rangle }{\left\langle min(T,R)\right\rangle }\,.\label{9}
\end{equation}
A probabilistic interpretation of this result and discussion with
examples are given in the SI.
\begin{figure}
\noindent \begin{centering}
\includegraphics[scale=0.32]{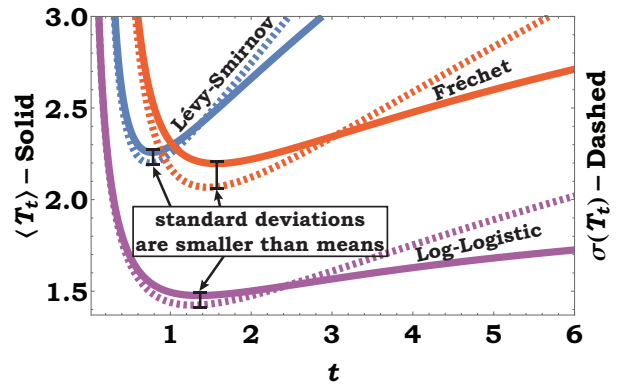}
\par\end{centering}
\caption{The mean (solid) and standard deviation (dashed) of the restarted
FPT $T_{t}$ vs. the sharp restart time $t$, for various distributions
of the underlying FPT $T$ (see SI for details). }
\end{figure}

\textbf{Conclusions and outlook. }In this letter we developed a theoretical
framework for first passage under restart. With its aid, we showed
how simple observations made for diffusion under restart can be elevated
to the level of generic statements which capture fundamental aspects
of the phenomena. The universal dominance of sharp restart over other
restart strategies is noteworthy. However, while this strategy can
be readily applied in some settings, its realization in others may
require going to extremes. Particularly, in biophysical settings the
generation of tight time distributions relies on the concatenation
of irreversible molecular transitions. Restart plays a role in such
systems \cite{Restart-Biophysics1,Restart-Biophysics2}, but the energetic
cost associated with creating an (almost) irreversible transition,
and the infinitely many required for \emph{mathematically} \emph{sharp}
restart, would surely give rise to interesting trade-offs. The incorporation
of such thermodynamic considerations into the framework presented
herein à la \cite{Barato1,Barato2}, and the identification of those
nearly optimal strategies (non-sharp but punctual) \cite{Punctual}
which perform best under energy consumption constraints, is yet a
future challenge. 

\textbf{Acknowledgments.} Shlomi Reuveni gratefully acknowledges support
from the James S. McDonnell Foundation via its postdoctoral fellowship
in studying complex systems. Shlomi Reuveni would like to thank Johan
Paulsson and the Department of Systems Biology at Harvard Medical
School for their hospitality. Arnab Pal acknowledges Saar Rahav for
many fruitful discussions.

\end{document}

\noindent \begin{center}
\textbf{\huge{}\uline{Supplementary Information}}\textbf{\LARGE{}}\\
\textbf{\LARGE{} }\\
\textbf{\LARGE{}}\\
\textbf{\LARGE{}}\\
\textbf{\LARGE{}}\\
\textbf{\Large{}First Passage Under Restart}\\
\textbf{\Large{}}\\
\textbf{\LARGE{}}\\
\textbf{\LARGE{}}\\
Arnab Pal$^{*}$~\&~Shlomi Reuveni$^{\dagger}$\\
~
\par\end{center}

\noindent \begin{center}
\textit{$^{*}$Schulich Faculty of Chemistry, Technion—Israel Institute
of Technology, Technion City, Haifa 32000, Israel.}
\par\end{center}

\noindent \begin{center}
\textit{$^{\dagger}$Department of Systems Biology, Harvard Medical
School, Boston, Massachusetts 02115, USA.}
\par\end{center}

\clearpage{}\tableofcontents{}\clearpage{}

\section{Derivation of Eq. (4) in main text}

The same line of argumentation brought in the text could also be used
to show that the sharp restart strategy, $Pr(R=t^{*})=1$, is optimal
among all possible stochastic restart strategies. Indeed, the law
of total expectation asserts that
\begin{equation}
\left\langle min(T,R)\right\rangle =\left\langle \left\langle min(T,R)|R\right\rangle _{T}\right\rangle _{R}\,,\label{SI1}
\end{equation}
where $\left\langle \cdot\right\rangle _{R}$ and $\left\langle \cdot\right\rangle _{T}$
explicitly mark the expectation with respect to $R$ and $T$. Utilizing
Eq. (3) in the text, and letting $I\{T<R\}=1-I\{T\geq R\}$ denote
an indicator random variable which takes the value one when $T<R$
and zero otherwise, it follows that
\begin{equation}
\begin{array}{l}
\left\langle T_{R}\right\rangle =\frac{\left\langle \left\langle min(T,R)|R\right\rangle _{T}\right\rangle _{R}}{Pr(T<R)}=\frac{\left\langle \frac{Pr(T<R|R)}{Pr(T<R|R)}\left\langle min(T,R)|R\right\rangle _{T}\right\rangle _{R}}{\left\langle I\{T<R\}\right\rangle _{R,T}}\\
\\
=\left\langle \frac{Pr(T<R|R)}{\left\langle I\{T<R\}\right\rangle _{R,T}}\frac{\left\langle min(T,R)|R\right\rangle _{T}}{Pr(T<R|R)}\right\rangle _{R}\,.\\
\text{ }
\end{array}\label{SI2}
\end{equation}
 However, note that since
\begin{equation}
\begin{array}{l}
\left\langle \frac{Pr(T<R|R)}{\left\langle I\{T<R\}\right\rangle _{R,T}}\right\rangle _{R}=\frac{\left\langle \left\langle I\{T<R\}|R\right\rangle _{T}\right\rangle _{R}}{\left\langle I\{T<R\}\right\rangle _{R,T}}=1\,,\\
\text{ }
\end{array}\label{SI3}
\end{equation}
 if an optimal sharp restart time $R=t^{*}$ exists, i.e., one which
satisfies
\begin{equation}
\begin{array}{l}
\frac{\left\langle min(T,t^{*})\right\rangle }{Pr(T<t^{*})}=\frac{\left\langle min(T,R)|R=t^{*}\right\rangle _{T}}{Pr(T<R|R=t^{*})}\\
\\
\leq\frac{\left\langle min(T,R)|R=t\right\rangle _{T}}{Pr(T<R|R=t)}=\frac{\left\langle min(T,t)\right\rangle }{Pr(T<t)}\,,\\
\text{ }
\end{array}\label{SI4}
\end{equation}
for all $t>0,$ then using Eqs. (\ref{SI2}) and (\ref{SI4}) we have
\begin{equation}
\frac{\left\langle min(T,t^{*})\right\rangle }{Pr(T<t^{*})}\leq\frac{\left\langle min(T,R)\right\rangle }{Pr(T<R)}=\left\langle T_{R}\right\rangle \,,\label{SI5}
\end{equation}
for every stochastic restart time $R$ and regardless of its distribution.

\section{Derivation of Eq. (5) in main text}

To derive Eq. (5) in the main text we note that 
\begin{equation}
\begin{array}{l}
\tilde{T}_{R}(s)=\left\langle e^{-sT_{R}}\right\rangle =Pr(T<R)\left\langle e^{-sT_{R}}|T<R\right\rangle \\
\\
+Pr(R\leq T)\left\langle e^{-sT_{R}}|R\leq T\right\rangle \,,\\
\text{ }
\end{array}\label{SI6}
\end{equation}
which gives 
\begin{equation}
\begin{array}{l}
\tilde{T}_{R}(s)=Pr(T<R)\left\langle e^{-s\{T_{R}\,|\,T<R\}}\right\rangle \\
\\
+Pr(R\leq T)\left\langle e^{-s\{T_{R}\,|\,R\leq T\}}\right\rangle \,.\\
\text{ }
\end{array}\label{SI7}
\end{equation}
However, utilizing Eq. (1) in the main text and recalling the way
$R_{min}$ and $T_{min}$ were defined, we see that
\begin{equation}
\begin{array}{l}
\{T_{R}\,|\,T<R\}=\{T\,|\,T<R\}\\
\\
=\{T\,|\,T=min(R,T)\}=T_{min}\,,\\
\text{ }
\end{array}\label{SI8}
\end{equation}
and
\begin{equation}
\begin{array}{l}
\{T_{R}\,|\,R\leq T\}=\{R+T_{R}^{\prime}\,|\,R\leq T\}\\
\\
=\{R\,|\,R=min(R,T)\}+T_{R}^{\prime}=R_{min}+T_{R}^{\prime}\,,\\
\text{ }
\end{array}\label{SI9}
\end{equation}
where in the second transition in Eq. (\ref{SI9}) we have further
used the fact that $T_{R}^{\prime}$ is an independent and identically
distributed copy of $T_{R}$ and hence independent of both $R$ and
$T$. We thus have 
\begin{equation}
\begin{array}{l}
\tilde{T}_{R}(s)=Pr(T<R)\left\langle e^{-sT_{min}}\right\rangle \\
\\
+Pr(R\leq T)\left\langle e^{-s\left(R_{min}+T_{R}^{\prime}\right)}\right\rangle \\
\\
=Pr(T<R)\tilde{T}_{min}(s)\\
\\
+Pr(R\leq T)\tilde{R}_{min}(s)\tilde{T}_{R}(s)\,,\\
\text{ }
\end{array}\label{SI10}
\end{equation}
where in the last step we have again used the fact that $T_{R}^{\prime}$
is an independent and identically distributed copy of $T_{R}$ and
hence $\left\langle e^{-s\left(R_{min}+T_{R}^{'}\right)}\right\rangle =\left\langle e^{-sR_{min}}\right\rangle \left\langle e^{-sT_{R}^{'}}\right\rangle =\left\langle e^{-sR_{min}}\right\rangle \left\langle e^{-sT_{R}}\right\rangle $.
Rearranging Eq. (\ref{SI10}) we have 
\begin{equation}
\tilde{T}_{R}(s)=\frac{Pr(T<R)\tilde{T}_{min}(s)}{1-Pr(R\leq T)\tilde{R}_{min}(s)}\,,\label{SI11}
\end{equation}
which identifies with Eq. (5) in the main text. 

\subsection{The case of exponential restart times}

When the restart time $R$ is exponentially distributed with rate
$r$ its probability density function is given by 
\begin{equation}
f_{R}(t)=re^{-rt}\,.\label{SI12}
\end{equation}
The terms in Eq. (\ref{SI11}) (Eq. (5) in the main text) can then
be worked out to give 
\begin{equation}
\begin{array}{l}
Pr(T<R)\tilde{T}_{min}(s)=Pr(T<R)\left\langle e^{-s\{T|T<R\}}\right\rangle \\
\\
=Pr(T<R)\frac{\left\langle \int_{T}^{\infty}f_{R}(t)e^{-sT}dt\right\rangle _{T}}{Pr(T<R)}=\left\langle e^{-sT}\int_{T}^{\infty}re^{-rt}dt\right\rangle _{T}\\
\\
=\left\langle e^{-sT}e^{-rT}\right\rangle _{T}=\left\langle e^{-(s+r)T}\right\rangle _{T}=\tilde{T}(r+s)\,,\\
\text{ }
\end{array}\label{SI13}
\end{equation}
and
\begin{equation}
\begin{array}{l}
Pr(R\leq T)\tilde{R}_{min}(s)=Pr(R\leq T)\left\langle e^{-s\{R|R\leq T\}}\right\rangle \\
\\
=Pr(R\leq T)\frac{\left\langle \int_{0}^{T}f_{R}(t)e^{-st}dt\right\rangle _{T}}{Pr(R\leq T)}=\left\langle \int_{0}^{T}re^{-rt}e^{-st}dt\right\rangle _{T}\\
\\
=\frac{r}{r+s}\left\langle 1-e^{-(s+r)T}\right\rangle _{T}=\frac{r}{r+s}\left(1-\tilde{T}(r+s)\right)\,.\\
\text{ }
\end{array}\label{SI14}
\end{equation}
Substituting back we find
\begin{equation}
\tilde{T}_{R}(s)=\frac{\tilde{T}(r+s)}{1-\frac{r}{r+s}\left(1-\tilde{T}(r+s)\right)}=\frac{\tilde{T}(r+s)}{\frac{s}{r+s}+\frac{r}{r+s}\tilde{T}(r+s)}\,,\label{SI15}
\end{equation}
which identifies with the result in the main text.

\section{Derivation of Eq. (7) in main text}

To derive Eq. (7) in the main text, we first recall the relation between
a Laplace transform of a random variable and its $n-th$ moment 
\begin{equation}
\left\langle Z^{n}\right\rangle =\left(-1\right)^{n}\frac{d^{n}\tilde{Z}(s)}{ds^{n}}|_{s=0}\,.\label{SI16}
\end{equation}
Recalling Eq. (\ref{SI10}) above
\begin{equation}
\tilde{T}_{R}(s)=Pr(T<R)\langle e^{-sT_{min}}\rangle+Pr(R\leq T)\langle e^{-s(R_{min}+T_{R}^{\prime})}\rangle\,,\label{SI17}
\end{equation}
we multiply both sides by -1, take a single derivative of with respect
to $s$, and the limit of $s\to0$, to obtain 
\begin{equation}
\begin{array}{l}
\langle T_{R}\rangle=Pr(T<R)\langle T_{min}\rangle+Pr(R\leq T)\langle R_{min}+T_{R}^{\prime}\rangle\\
\\
=Pr(T<R)\langle T_{min}\rangle+Pr(R\leq T)\left(\left\langle R_{min}\right\rangle +\left\langle T_{R}^{\prime}\right\rangle \right)\,.\\
\text{ }
\end{array}\label{SI18}
\end{equation}
This result is equivalent to Eq. (3) in the main text. Taking two
derivatives of Eq. (\ref{SI17}) with respect to $s$ and the limit
$s\to0$, we find
\begin{equation}
\begin{array}{l}
\langle T_{R}^{2}\rangle=Pr(T<R)\langle T_{min}^{2}\rangle+Pr(R\leq T)\langle(R_{min}+T_{R}^{\prime})^{2}\rangle\\
\\
=Pr(T<R)\langle T_{min}^{2}\rangle+Pr(R\leq T)[\langle R_{min}^{2}\rangle+\langle\left(T_{R}^{\prime}\right)^{2}\rangle+2\langle R_{min}T_{R}^{\prime}\rangle]\\
\\
=Pr(T<R)\langle T_{min}^{2}\rangle+Pr(R\leq T)[\langle R_{min}^{2}\rangle+\langle T_{R}^{2}\rangle+2\langle R_{min}\rangle\langle T_{R}\rangle]\,.\\
\text{ }
\end{array}\label{SI19}
\end{equation}
Noting that
\begin{equation}
\begin{array}{l}
\langle min(T,R)^{2}\rangle=Pr(T<R)\langle T^{2}|T<R\rangle+Pr(R\leq T)\langle R^{2}|R\leq T\rangle\\
\\
=Pr(T<R)\langle T_{min}^{2}\rangle+Pr(R\leq T)\langle R_{min}^{2}\rangle\,,\\
\text{ }
\end{array}\label{SI20}
\end{equation}
 using Eq. (3) in the main text, and rearranging terms in Eq. (\ref{SI19}),
we recover Eq. (7) in the main text
\begin{equation}
\langle T_{R}^{2}\rangle=\frac{\langle min(T,R)^{2}\rangle}{Pr(T<R)}+\frac{2Pr(R\leq T)\langle R_{min}\rangle\langle min(T,R)\rangle}{Pr(T<R)^{2}}\,.\label{SI21}
\end{equation}

An equivalent way in which Eq. (7) can be derived is by recalling
that the Laplace transform of a random variable has the following
moment expansion 
\begin{equation}
\tilde{Z}(s)=\left\langle e^{-sZ}\right\rangle =1-s\left\langle Z\right\rangle +\frac{1}{2}s^{2}\left\langle Z^{2}\right\rangle +o(s^{2})\,.\label{SI24}
\end{equation}
On the one hand we thus have 
\begin{equation}
\tilde{T}_{R}(s)=1-\left\langle T_{R}\right\rangle s+\frac{1}{2}\left\langle T_{R}^{2}\right\rangle s^{2}+o(s^{2}),\label{SI25}
\end{equation}
and from the other by use of Eq. (5) in the main text 
\begin{equation}
\tilde{T}_{R}(s)=\frac{Pr(T<R)\left[1-s\left\langle T_{min}\right\rangle +\frac{1}{2}s^{2}\left\langle T_{min}^{2}\right\rangle +o(s^{2})\right]}{1-Pr(R\leq T)\left[1-s\left\langle R_{min}\right\rangle +\frac{1}{2}s^{2}\left\langle R_{min}^{2}\right\rangle +o(s^{2})\right]}\,.\label{SI26}
\end{equation}
Expanding the right hand side of Eq. (\ref{SI26}) to second order
in the Laplace variable ``$s$'' and equating coefficients of equal
powers we find 
\begin{equation}
\left\langle T_{R}\right\rangle =\frac{Pr(R\leq T)\left\langle R_{min}\right\rangle +Pr(T<R)\left\langle T_{min}\right\rangle }{Pr(T<R)}=\frac{\left\langle min(T,R)\right\rangle }{Pr(T<R)}\,,\label{SI27}
\end{equation}
and
\begin{equation}
\begin{array}{l}
\left\langle T_{R}^{2}\right\rangle =2\left[\left\langle R_{min}\right\rangle ^{2}+\frac{\left\langle R_{min}\right\rangle ^{2}}{Pr(T<R)^{2}}-\frac{2\left\langle R_{min}\right\rangle ^{2}}{Pr(T<R)}-\left\langle R_{min}\right\rangle \left\langle T_{min}\right\rangle \right.\\
\\
\left.+\frac{\left\langle R_{min}\right\rangle \left\langle T_{min}\right\rangle }{Pr(T<R)}-\frac{\left\langle R_{min}^{2}\right\rangle }{2}+\frac{\left\langle R_{min}^{2}\right\rangle }{2Pr(T<R)}+\frac{\left\langle T_{min}^{2}\right\rangle }{2}\right]\\
\\
=2\left[\left\langle R_{min}\right\rangle ^{2}\frac{Pr(T\geq R)^{2}}{Pr(T<R)^{2}}+\left\langle R_{min}\right\rangle \left\langle T_{min}\right\rangle \frac{Pr(T\geq R)}{Pr(T<R)}\right.\\
\\
\left.+\left\langle R_{min}^{2}\right\rangle \frac{Pr(T\geq R)}{2Pr(T<R)}+\frac{\left\langle T_{min}^{2}\right\rangle }{2}\right]\\
\text{ }
\end{array}\label{SI28}
\end{equation}
Equation (\ref{SI27}) reaffirms Eq. (3) in the main text, rearranging
Eq. (\ref{SI28}) we find
\begin{equation}
\begin{array}{l}
\left\langle T_{R}^{2}\right\rangle =2\left[\frac{\left\langle R_{min}^{2}\right\rangle Pr(T\geq R)+\left\langle T_{min}^{2}\right\rangle Pr(T<R)}{2Pr(T<R)}\right.\\
\\
\left.+\frac{\left\langle R_{min}\right\rangle Pr(T\geq R)\left(\left\langle R_{min}\right\rangle Pr(T\geq R)+\left\langle T_{min}\right\rangle Pr(T<R)\right)}{Pr(T<R)^{2}}\right]\\
\\
=\frac{\left\langle min(T,R)^{2}\right\rangle }{Pr(T<R)}+\frac{2Pr(T\geq R)\left\langle R_{min}\right\rangle \left\langle min(T,R)\right\rangle }{Pr(T<R)^{2}}\,,\\
\text{ }
\end{array}\label{SI29}
\end{equation}
which coincides with Eq. (7) in the main text.

\section{\label{sec:Derivation-of-Eq. 8}Derivation of Eq. (8) in main text}

To derive Eq. (8) in the main text we first show that when a process
with first passage time $T$, such that $\sigma\left(T\right)/\left\langle T\right\rangle >1$,
is restarted at some low constant rate $r$, the mean first passage
time of the restarted process will surely be smaller than $\left\langle T\right\rangle $.
We start by multiplying both sides of Eq. (\ref{SI15}) by -1, taking
a single derivative with respect to $s$, and the limit of $s\to0$
to obtain\footnote{Alternatively, Eq. (\ref{SI30}) could also be obtained directly from
from Eq. (3) in the main text by exploiting the fact that when the
restart time $R$ is exponential with rate $r$ its probability density
function is given by $f_{R}(t)=re^{-rt}.$} 
\begin{equation}
\left\langle T_{R}\right\rangle =\left(1-\tilde{T}(r)\right)/\left(r\tilde{T}(r)\right)\,.\label{SI30}
\end{equation}
To probe the behavior of $\left\langle T_{R}\right\rangle $ in the
limit $r\rightarrow0$, we expand $\tilde{T}(r)$ to second order
\begin{equation}
\tilde{T}(r)=1-\left\langle T\right\rangle r+\frac{1}{2}\left\langle T^{2}\right\rangle r^{2}+o(r^{2})\,,\label{SI31}
\end{equation}
and substitute the result back into Eq. (\ref{SI30}) to give
\begin{equation}
\left\langle T_{R}\right\rangle =\frac{\left\langle T\right\rangle r-\frac{1}{2}\left\langle T^{2}\right\rangle r^{2}-o(r^{2})}{r-\left\langle T\right\rangle r^{2}+\frac{1}{2}\left\langle T^{2}\right\rangle r^{3}+o(r^{3})}\,.\label{SI32}
\end{equation}
It is now easy to see that
\begin{equation}
\begin{array}{l}
\left\langle T_{R}\right\rangle =\left\langle T\right\rangle +\left[\left\langle T\right\rangle ^{2}-\frac{1}{2}\left\langle T^{2}\right\rangle \right]r+o(r^{2})\\
\\
=\left\langle T\right\rangle +\frac{1}{2}\left[\left\langle T\right\rangle ^{2}-\sigma^{2}\left(T\right)\right]r+o(r^{2})\,,\\
\text{ }
\end{array}\label{SI33}
\end{equation}
which in turn means that when $\sigma\left(T\right)/\left\langle T\right\rangle >1$
and when $r$ is sufficiently small 
\begin{equation}
\left\langle T_{R}\right\rangle <\left\langle T\right\rangle \,.\label{SI34}
\end{equation}
In deriving Eq. (\ref{SI34}) above we did not make any assumptions
with regard to the origin or distribution of the first passage time
$T$ albeit the assumption that $\sigma\left(T\right)/\left\langle T\right\rangle >1$.
In principle, $T$ could also be the first passage time of a process
which is already subject to restart and even that of a process which
is subject to \emph{optimal sharp restart}. However, and as explained
in the main text, the latter case is impossible as it would in turn
mean that a non-sharp, or mixed, restart strategy can attain lower
mean first passage times than those attained for optimal sharp restart. 

\section{Details of distributions in Fig. 5 }

Plot were made for the following distributions:
\begin{enumerate}
\item Lévy-Smirnov (FPT of diffusion mediated search)
\begin{equation}
f_{T}(t)=\sqrt{L^{2}/4D\pi t^{3}}e^{-L^{2}/4Dt}\,,\label{SI35}
\end{equation}
$(t\geq0)$, with $\sqrt{L^{2}/4D}=0.65$.
\item Fréchet
\begin{equation}
Pr(T\leq t)=e^{-t^{-\alpha}}\,,\label{SI36}
\end{equation}
 $(t\geq0)$, with $\alpha=1$. 
\item Log-Logistic
\begin{equation}
Pr(T\leq t)=\left[1+\left(t/\alpha\right)^{-\beta}\right]^{-1}\,,\label{SI37}
\end{equation}
 $(t\geq0)$, with $\alpha=1$ and $\beta=3/2$.
\end{enumerate}

\section{Derivation and probabilistic interpretation of Eq. (9) in main text}

To derive Eq. (9) in the main text we first note that
\begin{equation}
\sigma\left(T_{R}\right)/\left\langle T_{R}\right\rangle \leq1\Longleftrightarrow\left\langle T_{R}^{2}\right\rangle \leq2\left\langle T_{R}\right\rangle ^{2}\,.\label{SI38}
\end{equation}
Plugging in Eqs. (3) and (7) in the main text we write the right inequality
in (\ref{SI38}) explicitly 
\begin{equation}
\frac{\left\langle min(T,R)^{2}\right\rangle }{Pr(T<R)}+\frac{2Pr(R\leq T)\left\langle R_{min}\right\rangle \left\langle min(T,R)\right\rangle }{Pr(T<R)^{2}}\leq2\frac{\left\langle min(T,R)\right\rangle ^{2}}{Pr(T<R)^{2}}\,,\label{SI39}
\end{equation}
and rearrange to obtain 
\begin{equation}
\frac{\left\langle min(T,R)^{2}\right\rangle }{2}\leq\frac{\left\langle min(T,R)\right\rangle }{Pr(T<R)}\left[\left\langle min(T,R)\right\rangle -Pr(R\leq T)\left\langle R_{min}\right\rangle \right]\,.\label{SI40}
\end{equation}
Recalling that 
\begin{equation}
\left\langle min(T,R)\right\rangle =Pr(T<R)\left\langle T_{min}\right\rangle +Pr(R\leq T)\left\langle R_{min}\right\rangle \,,\label{SI41}
\end{equation}
we see that Eq. (\ref{SI40}) can be written as 
\begin{equation}
\left\langle T_{min}\right\rangle \geq\frac{1}{2}\frac{\left\langle min(T,R)^{2}\right\rangle }{\left\langle min(T,R)\right\rangle }\,,\label{SI42}
\end{equation}
which proves Eq. (9) in the main text.

The result in Eq. (9) could be understood probabilistically by considering
the effect the introduction of a low restart rate, $r\ll1$, has on
the mean FPT, $\left\langle T_{R}\right\rangle $, of a process that
is already subject to restart. In section (\ref{sec:Derivation-of-Eq. 8})
above we have shown that this would lower $\left\langle T_{R}\right\rangle $
if $\sigma\left(T_{R}\right)/\left\langle T_{R}\right\rangle >1$,
but would increase $\left\langle T_{R}\right\rangle $ (or leave it
unchanged) otherwise. Equation (9) in the main text asserts that we
can replace $\sigma\left(T_{R}\right)/\left\langle T_{R}\right\rangle >1$
in the previous sentence with $\frac{1}{2}\frac{\left\langle min(T,R)^{2}\right\rangle }{\left\langle min(T,R)\right\rangle }>\left\langle T_{min}\right\rangle $.
To better understand why, imagine that a low restart rate $r\ll1$
is introduced to the system from the outside and consider the effect
this will have on the expected completion time of the process. The
external restart rate added is infinitesimally small, but it will
eventually find the process at some random point in time and restart
it. The expected completion time from that point onward is $\left\langle T_{R}\right\rangle $
(to an excellent approximation) as an additional, exogenous, restart
event within this time frame is extremely unlikely. This expected
time to completion will now be compared to that which would have been
attained in the absence of exogenous restart. 

Immediately after our process has started the mean time taken for
it to either complete or restart is given by $\left\langle min(T,R)\right\rangle $.
However, when the process is visited at some random point in time
(after it has already started rolling) this is no longer the case.
Indeed, one is then interested in the mean \emph{residual time} (averaged
over random points in time) that is left until either restart or completion
happen, and for renewal processes this time is generally known to
be given by \cite{Gallager}
\begin{equation}
\left\langle T_{res}\right\rangle =\frac{1}{2}\frac{\left\langle min(T,R)^{2}\right\rangle }{\left\langle min(T,R)\right\rangle }\,.\label{SI43}
\end{equation}
And so, in the absence of exogenous restart, one would need to wait
$\left\langle T_{res}\right\rangle $ units of time (on average) for
the process to either complete or restart (counting from that random
point in time at which the process was visited). To that, one needs
not add a thing if the process completes, or is required to add $\left\langle T_{R}\right\rangle $
units of time (on average) if the processes restarts. What is, however,
the probability that the latter happens? 

To answer this, we once again observe that if we were to examine our
process immediately after it has started, the answer would have been
$Pr(R\leq T)$. However, we are now interested in the probability
that $R\leq T$ given that we observe the process at a random point
in time and after it has already been given the opportunity to ``age''.
This probability is in turn given by $\frac{Pr(R\leq T)\left\langle R_{min}\right\rangle }{\left\langle min(T,R)\right\rangle }=\frac{Pr(R\leq T)\left\langle R_{min}\right\rangle }{Pr(T<R)\left\langle T_{min}\right\rangle +Pr(R\leq T)\left\langle R_{min}\right\rangle }$
which is the exact relative fraction, on the time axis, captured by
those time spans which end with (endogenous) restart rather than completion.
We thus find that the introduction of a low restart rate will increase
the mean FPT (or leave it unchanged) whenever 
\begin{equation}
\left\langle T_{R}\right\rangle \geq\left\langle T_{res}\right\rangle +\frac{Pr(R\leq T)\left\langle R_{min}\right\rangle }{\left\langle min(T,R)\right\rangle }\left\langle T_{R}\right\rangle \,,\label{SI44}
\end{equation}
and would otherwise lower it. Plugging in the expressions for $\left\langle T_{res}\right\rangle $
and $\left\langle T_{R}\right\rangle $ into Eq. (\ref{SI44}) we
have 
\begin{equation}
\frac{\left\langle min(T,R)\right\rangle }{Pr(T<R)}\geq\frac{1}{2}\frac{\left\langle min(T,R)^{2}\right\rangle }{\left\langle min(T,R)\right\rangle }+\frac{Pr(R\leq T)\left\langle R_{min}\right\rangle }{Pr(T<R)}\,.\label{SI45}
\end{equation}
and rearrangement then gives back Eq. (\ref{SI42}).

\section{Numerical exploration of Eq. (9) in the main text}

To demonstrate the validity and generality of Eq. (9) in the main
text we explore several numerical examples. In Figure (S1), we revisit
the examples given in Fig. 5 (main text). There, the restart time
distribution was sharp ($Pr(R=t)=1$), and we now show that $\sigma\left(T_{R}\right)/\left\langle T_{R}\right\rangle \leq1\Longleftrightarrow\left\langle T_{min}\right\rangle \geq\frac{1}{2}\frac{\left\langle min(T,R)^{2}\right\rangle }{\left\langle min(T,R)\right\rangle }$
for every $t\geq0$ (regardless of its optimality). We do this to
emphasize that Eq. (9) in the main text generalizes Eq. (8) to sharp,
but not necessarily optimal, restart times.
\begin{figure}[H]
\noindent \begin{centering}
\includegraphics[scale=0.6]{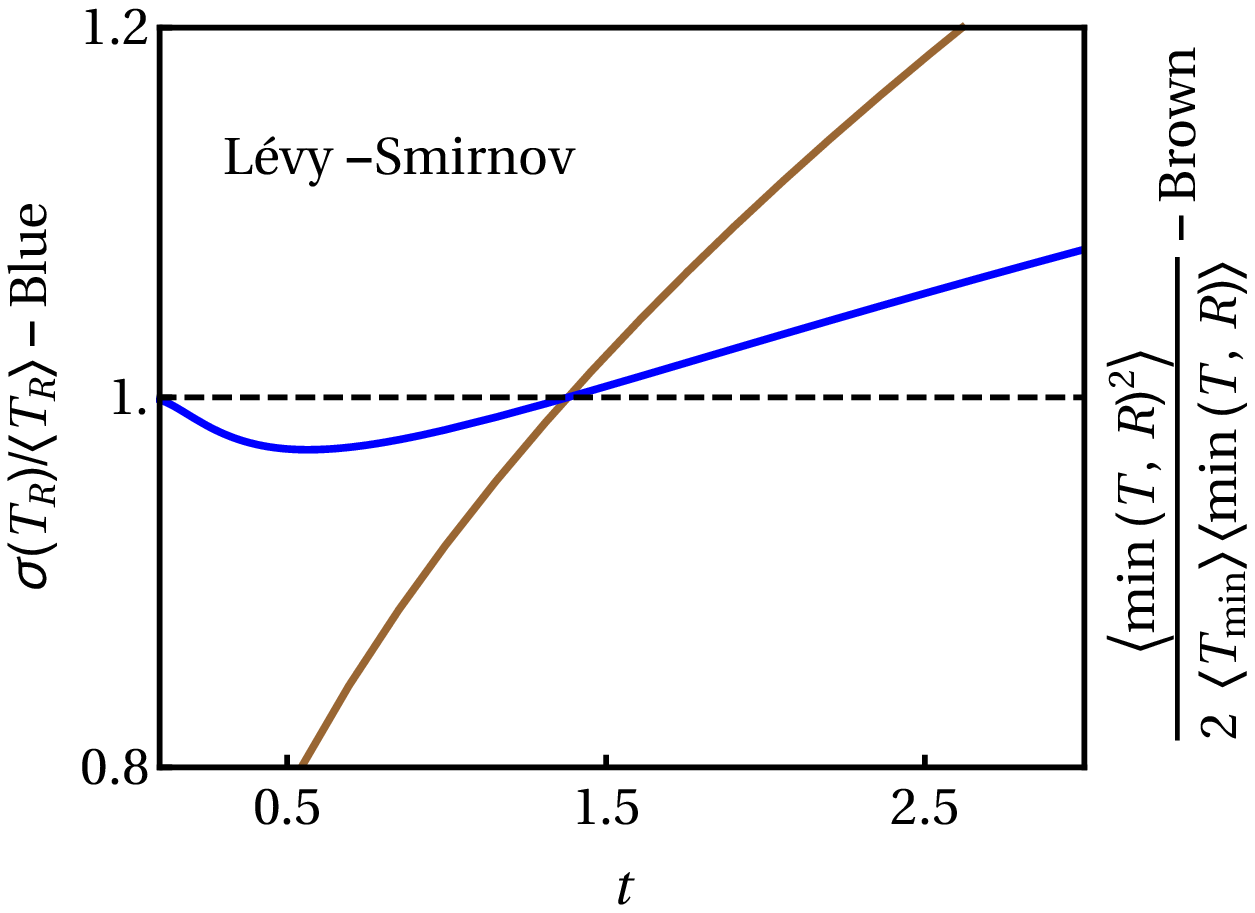}
\par\end{centering}
\noindent \begin{centering}
\includegraphics[scale=0.6]{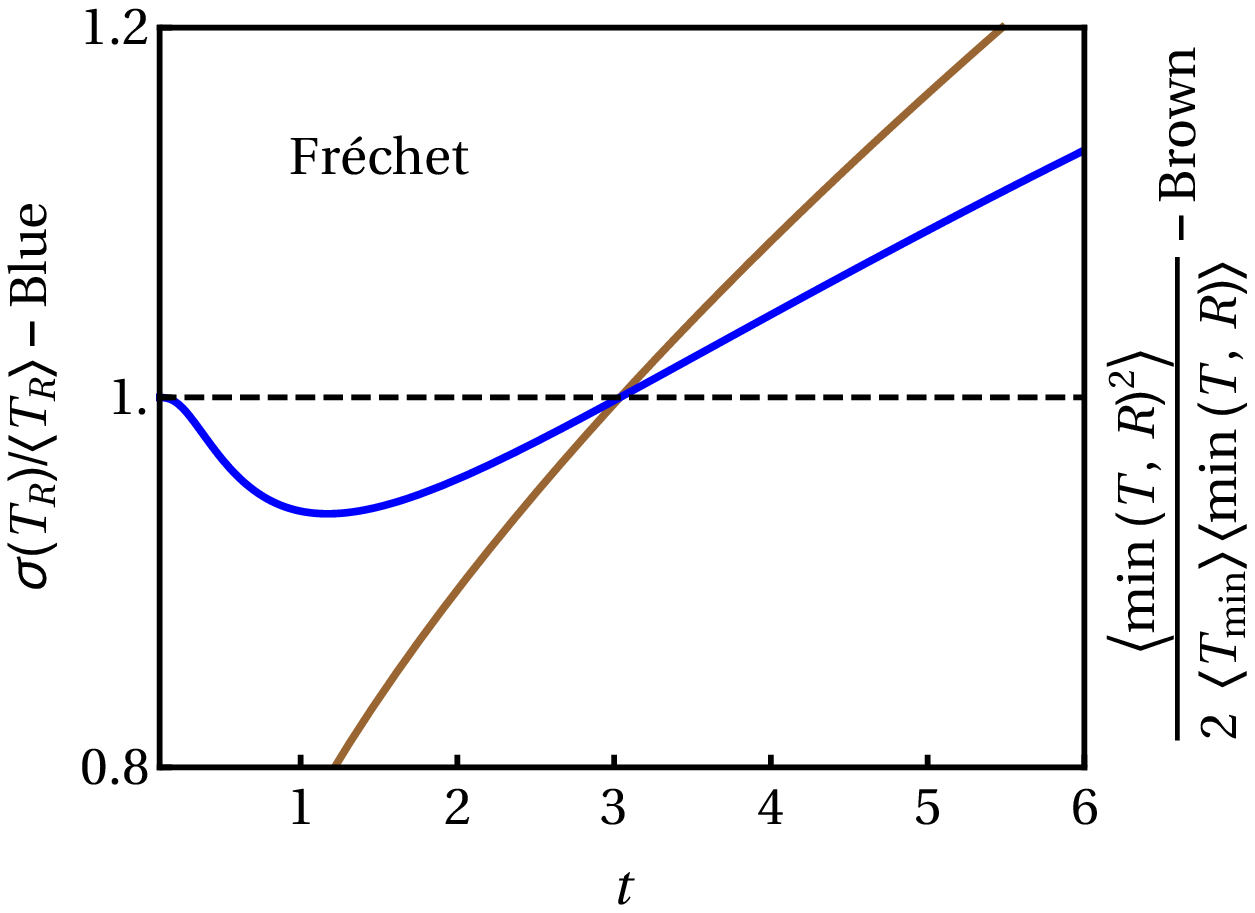} 
\par\end{centering}
\noindent \begin{centering}
\includegraphics[scale=0.6]{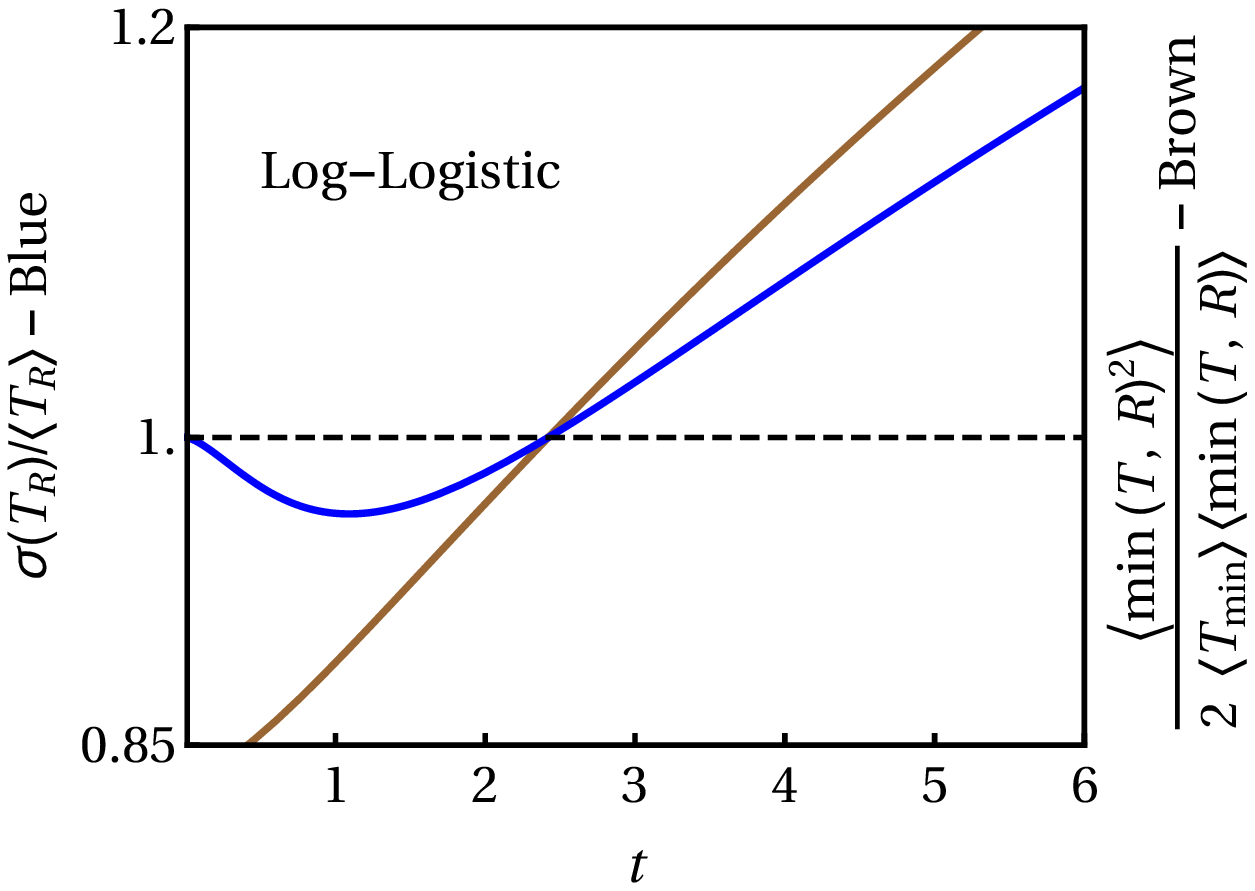}
\par\end{centering}
\noindent Figure S1: Plots demonstrating that Eq. (9) in the main
text holds for all three cases considered in Fig. (5). From top to
bottom: Lévy-Smirnov, Fréchet, Log-Logistic.
\end{figure}

In Figure (S2), we take $T$ as in Eq. (\ref{SI35}) above and examine
three different restart time distributions (Uniform, Gamma, Weibull)
for $R$. We show that $\sigma\left(T_{R}\right)/\left\langle T_{R}\right\rangle \leq1\Longleftrightarrow\left\langle T_{min}\right\rangle \geq\frac{1}{2}\frac{\left\langle min(T,R)^{2}\right\rangle }{\left\langle min(T,R)\right\rangle }$
in all three cases and regardless of the optimality of mean restart
time $\left\langle R\right\rangle $. We do this to emphasize that
Eq. (9) in the main text generalizes Eq. (8) to any restart time distribution.
\begin{figure}[H]
\noindent \begin{centering}
\includegraphics[scale=0.6]{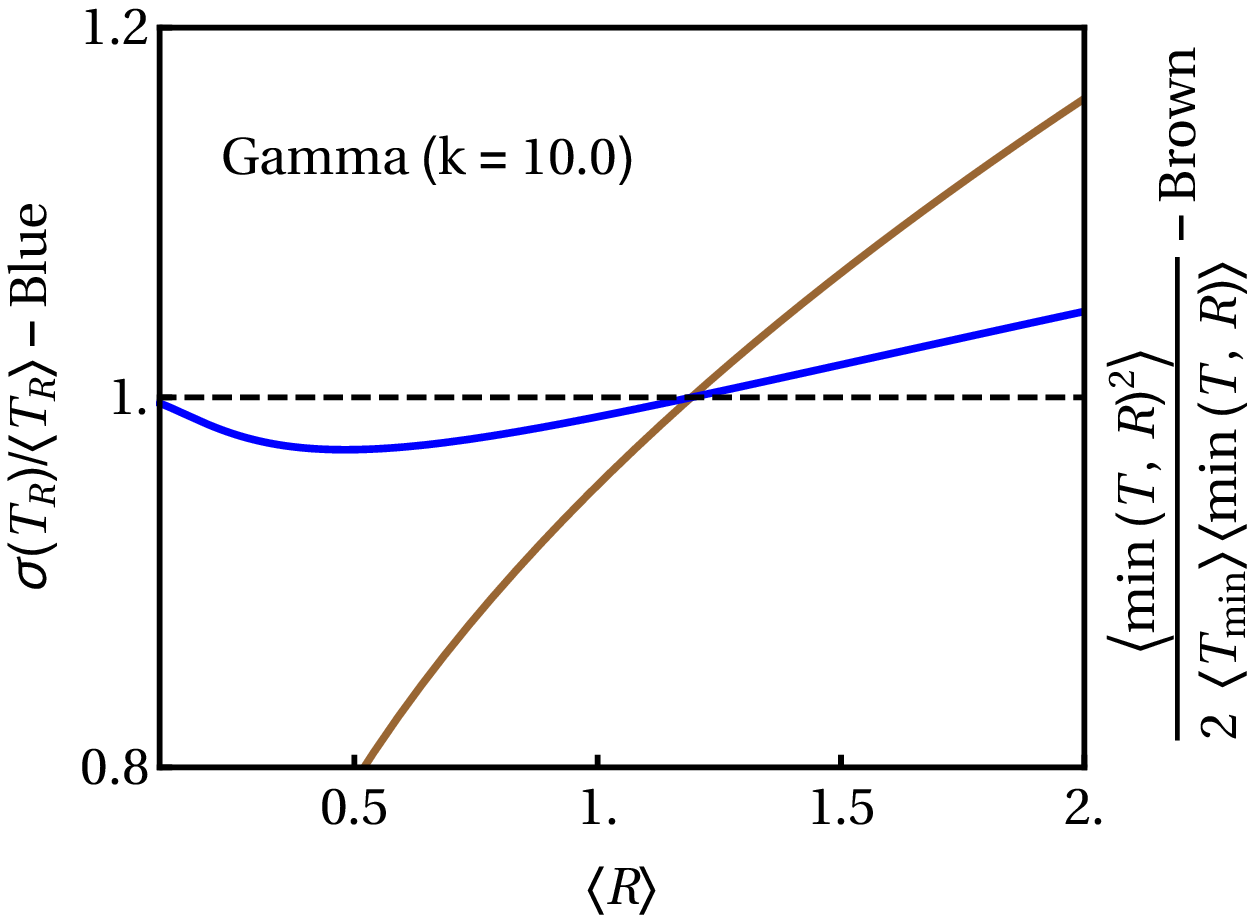}
\par\end{centering}
\noindent \begin{centering}
\includegraphics[scale=0.6]{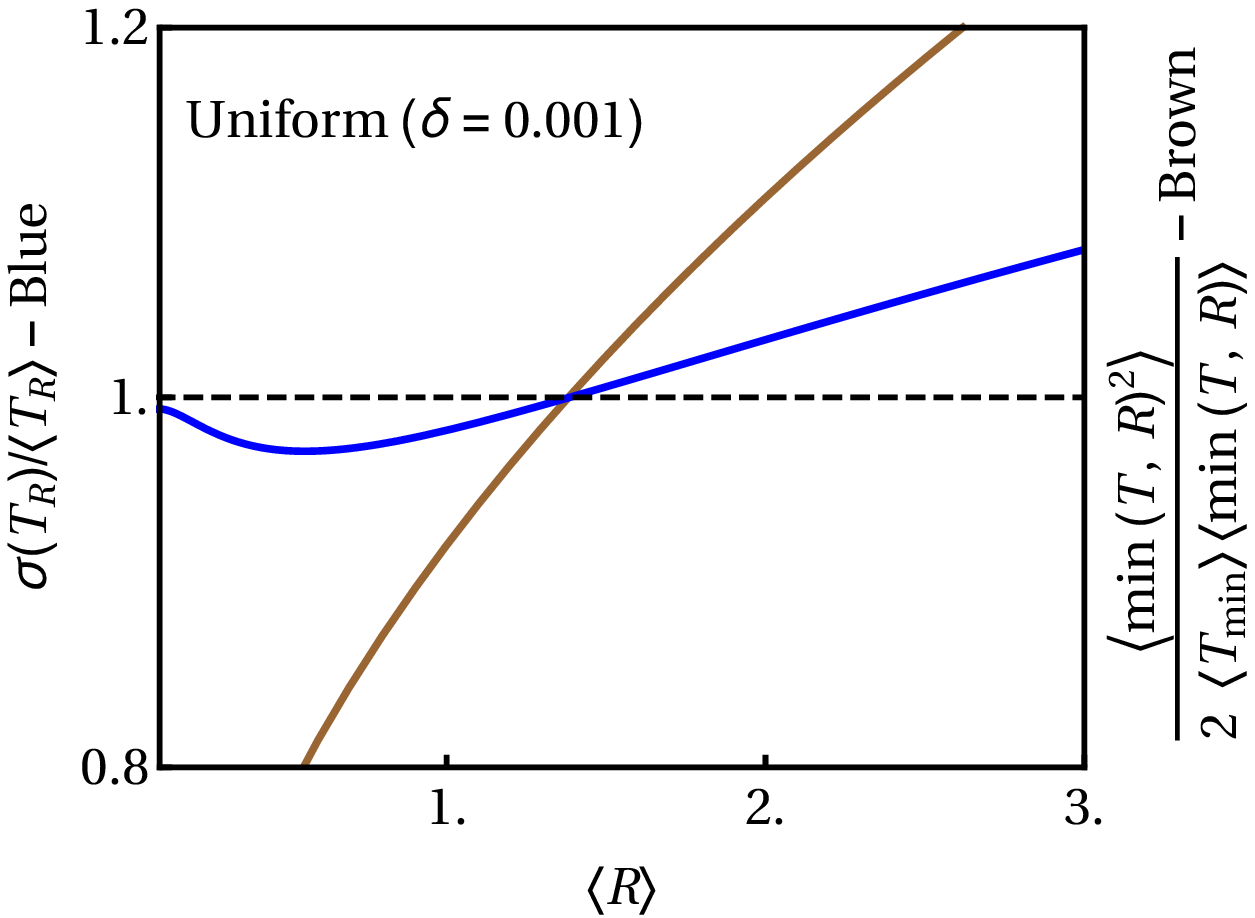}
\par\end{centering}
\noindent \begin{centering}
\includegraphics[scale=0.6]{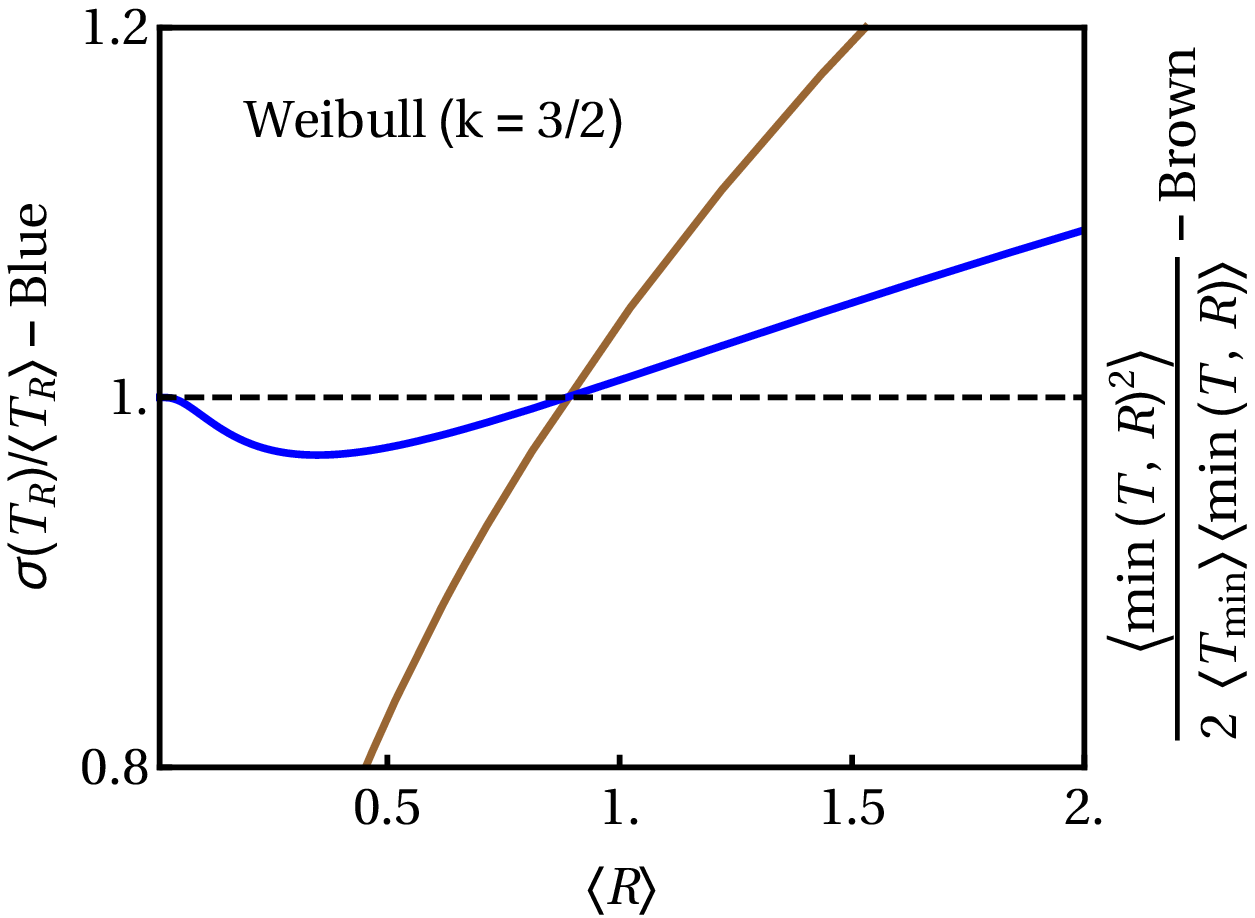}
\par\end{centering}
\noindent Figure S2: Plots demonstrating that Eq. (9) in the main
text holds for restart time distributions other than the sharp. From
top to bottom: (i) Gamma distribution $f_{R}(t)=\frac{1}{\Gamma(k)\theta^{k}}t^{k-1}e^{-t/\theta}$
with $k=10$ and $\theta=\left\langle R\right\rangle /k$; (ii) Uniform
distribution $f_{R}(t)=1/\delta$ for $t\in[\left\langle R\right\rangle -\delta/2,\left\langle R\right\rangle +\delta/2]$
and $f_{R}(t)=0$ otherwise; (iii) Weibull distribution $f_{R}(t)=\frac{k}{\lambda}\left(\frac{t}{\lambda}\right)^{k-1}e^{-\left(t/\lambda\right)^{k}}$
with $k=3/2$ and $\lambda=\left\langle R\right\rangle /\Gamma(5/3)$. 
\end{figure}